\documentclass[preprint]{aastex}

\shortauthors{Joo \& Lee}
\shorttitle{SFHs of GCs with Multiple Pops. I.}

\begin{document}

\title{STAR FORMATION HISTORIES OF GLOBULAR CLUSTERS WITH MULTIPLE POPULATIONS. I. $\omega$~CEN, M22, AND NGC~1851}

\author{
Seok-Joo Joo and
Young-Wook Lee\altaffilmark{1}
}

\affil{Center for Galaxy Evolution Research and Department of Astronomy, Yonsei University, Seoul 120-749, Korea}

\altaffiltext{1}{To whom correspondence should be addressed. E-mail: ywlee2@yonsei.ac.kr}

\begin{abstract}

There is increasing evidence that some massive globular clusters (GCs) host multiple stellar populations
having different heavy element abundances enriched by supernovae. They usually accompany multiple
red giant branches (RGBs) in the color-magnitude diagrams (CMDs), and are distinguished from most of the other GCs
which display variations only in light element abundances. In order to investigate the star formation histories
of these peculiar GCs, we have constructed synthetic CMDs for $\omega$~Cen, M22, and NGC~1851.
Our models are based on the updated versions of Yonsei-Yale (Y$^2$) isochrones and horizontal branch (HB)
evolutionary tracks which include the cases of enhancements in both helium and the total CNO abundances.
To estimate ages and helium abundances of subpopulations in each GC, we have compared our models with
the observations on the Hess diagram by employing a $\chi^2$ minimization technique.
We find that metal-rich subpopulations in each of these GCs are
also enhanced in helium abundance, and the age differences between the metal-rich and metal-poor subpopulations
are fairly small ($\sim$0.3$-$1.7 Gyr), even in the models with the observed variations in the total
CNO content. These are required to simultaneously reproduce the observed extended HB and the splits
on the main sequence, subgiant branch, and RGB.
Our results are consistent with the hypothesis that these GCs are the relics of more massive primeval dwarf galaxies
that merged and disrupted to form the proto-Galaxy.

\end{abstract}


\section{INTRODUCTION}

During the past decade, new observations have drastically changed our conventional view on the stellar population
content of globular clusters (GCs) in the Milky Way. It is now generally believed that
most GCs possess two or more stellar populations rather than a single population having the same age
and homogeneous chemical composition. Photometric observations have revealed clear signatures of
multiple populations in many of the massive GCs, in all branches on the color-magnitude diagram
\citep[CMD; e.g.,][]{lee99,pan00,bed04,sie07,pio07,mil08,mar08,pio09,mor09,jwlee09a,fer09,han09,roh11}.
Spectroscopic studies have established star-to-star variations in light element abundances such as C, N, O,
Na, Mg, and/or Al (\citealp[e.g.,][and references therein]{car09a,car09b}; \citealp[see also][for a recent review]{gra12a}).
Large variations of helium abundance are also expected in many GCs \citep[e.g.,][]{dan04,nor04,lee05,lee07,
pio05,pio07,ren08,yoo08,dup11,mon11,moe11,gra12b}. As proposed by \citet{car10b} and \citet{mar11b}, now ``$normal$" GCs
can be defined, in terms of chemical composition, as those that have homogeneous abundance in heavy elements,
such as calcium and iron, but show variations in light elements such as a Na-O anticorrelation.
The presence of multiple populations and chemical inhomogeneity in these GCs is mainly considered
to be due to the pollution from the fast-rotating massive stars \citep[FRMSs;][]{dec07} and/or intermediate mass
asymptotic giant branch (IMAGB) stars \citep{dan07,ven08}.

However, some massive GCs, such as $\omega$~Centauri \citep{lee99,bed04}, M54 \citep{lay97,sie07},
M22 \citep{mar09,dac09,jwlee09a}, NGC~1851 \citep{han09,car10c}, Terzan~5 \citep{fer09},
and NGC~2419 \citep{coh10,dic11}, are distinguished from the normal GCs. They show evidence
of supernovae (SNe) enrichment in addition to the spreads in light elements.
The discrete distributions of red giant branches (RGBs) observed in the CMDs of these GCs,
together with spectroscopic variations in heavy element abundances, indicate successive formation of stellar
generations from the gas enriched by SNe. They are generally thought to be the relics
of more massive primeval dwarf galaxies that disrupted by and merged with the Galaxy \citep{lee99,gne02,bek03,rom07,bok08},
and therefore, have important implications on the hierarchical merging paradigm of galaxy formation.

The purpose of this paper is to investigate the star formation histories of three GCs that belong to the
latter category, i.e., $\omega$~Cen, M22, and NGC~1851. In order to estimate ages, metallicities,
and helium abundances of subpopulations in these GCs, we have constructed synthetic CMDs
which best reproduce the observations. The construction of synthetic CMDs were performed
in a self-consistent manner from main sequence (MS) to horizontal branch (HB), whose morphologies vary with age,
metallicity, and helium abundance. In the following section, we describe our construction of stellar population models.
The synthetic CMDs, together with our best estimates of model parameters, for the three GCs are given in Section 3.
Finally, based on our results, we briefly discuss possible formation scenarios of these GCs in Section 4.


\section{MODEL CONSTRUCTION}

\subsection{Stellar Population Models}

The synthetic CMDs were constructed following the techniques developed by \citet{lee90,lee94}
for the HB and RR Lyrae variable stars, and by \citet{par97} for the MS to RGB \citep[see also][]{lee05,yoo08}.
Our models are based on the latest version of Yonsei-Yale (Y$^2$) isochrones and
a new set of Y$^2$ HB evolutionary tracks for the normal and enhanced helium abundances, respectively
(\citealp{yi08}; Y.-W. Lee et al. 2012, in preparation).
The HB evolutionary tracks used in our modeling are the updated version of those presented by \citet{ld90}
and \citet{yi97}, and they have been calculated using the same code and input physics employed
in the construction of Y$^2$ isochrones. Recent observations have shown that [CNO/Fe] is enhanced
in metal-rich subpopulation in most of GCs being considered in this paper \citep[e.g.,][]{yon09,mar11b,mar12a,
mar12b,gra12b,alv12}. In order to reflect this in our modeling, we have expanded the parameter space
of Y$^2$ isochrones and HB evolutionary tracks to include the cases of normal and enhanced nitrogen
abundances ([N/Fe] = 0.0, 0.8, and 1.6; Y.-W. Lee et al. 2012, in preparation).
The observed variations in the total CNO abundance were reproduced by interpolating these nitrogen enhanced
isochrones and HB evolutionary tracks. Our test simulations with varying N and O abundances show that, once
the total CNO sum ([CNO/Fe]) is held constant, both N and O have almost identical effects on the HR diagram
\citep[see also][]{ven09,sbo11,van12}.
Model ingredients adopted in our construction of synthetic CMDs are summarized in Table 1.

In the evolution of low-mass stars, mass loss on the RGB plays an important role in determining
the morphology of HB. In order to estimate the amount of this mass-loss, following \citet{lee94} and \citet{rey01},
we have adopted the empirical formula of \citet{rei77} and calibrated its coefficient, $\eta$,
to the HB morphologies of old GCs in the inner halo (R$_{GC}$ $<$ 8 kpc).
We have first integrated mass-loss rate with time steps along the evolutionary tracks
from the low MS to the RGB tip, and tabulated the resulting values of mass-loss
as functions of $\eta$, age, and chemical composition. Using these tables, together with the values of
the RGB tip masses (calculated without the mass-loss) taken from the Y$^2$ isochrones,
we constructed a series of synthetic HB models to form HB isochrones.
Figure 1 compares these HB isochrones with the inner halo GCs on the HB morphology-metallicity diagram,
under different assumptions as to the value of $\eta$. Assuming that these inner halo GCs are coeval and
have a mean age of 13 Gyr \citep{maf09,dot10}, we obtained $\eta\approx$ 0.53, which best reproduces
the observed correlation. We have used this same value of $\eta$ in all of our model constructions.
For the populations enhanced in helium and CNO abundances, the same value of $\eta$ was adopted
under the assumption that the mass-loss efficiency parameter, $\eta$, would not be changed
with helium or CNO abundances. Note, however, that the changes in gravity and luminosity caused
by helium and CNO enhancements were still reflected in our calculation of the amount of mass-loss.
Note also that the empirical fitting factor, $\eta$, depends on the stellar libraries chosen and the
ages assumed. For example, $\eta$ would be increased to $\sim$0.59,
if we adopted 12 Gyr, instead of 13 Gyr, as the mean age for the inner halo GCs.

Theoretical HR diagrams (log $L$ vs. log $T \rm_{eff}$) of our models were converted into observable
quantities (magnitude vs. color) by employing the semiempirical color table of \citet{gre87} and the synthetic model
atmospheres of \citet{cas03}. While the \citet{gre87} table has been adopted to the $Y^2$ isochrones and proven to show
reasonable agreements with observed CMDs of GCs \citep{yi01,kim02}, it is only available for the Johnson-Cousins
$UBVRI$ system. Therefore, following the method described in \citet{yi95}, we have generated
synthetic color tables from the \citet{cas03} stellar spectral library for the Johnson-Cousins $UBVRI$
\citep{bes90a}, $Hubble~Space~Telescope$ ($HST$) ACS/WFC, $HST$ WFC3/UVIS, and
$Ca$-$uvby$ \citep{cra70,ant91} filter systems.
The bolometric corrections were computed using the definition of \citet{bes98}, and the magnitudes
and color indices were calibrated with the model flux of Vega \citep{cas03} to the relevant zeropoints
of each filter system (i.e., \citealt{bes90b} for the Johnson-Cousins $UBVRI$, \citealt{bed05} and \citealt{dot07}
for the $HST$ ACS/WFC and $HST$ WFC3/UVIS, and \citealt{cra70} for the Str\"{o}mgren $uvby$ filter systems).
The $hk$ index was separately normalized with the model flux of the standard star HD 83373, which has
the closest spectral type to that of Vega in the catalog of \citet{ant91}, assuming $T \rm_{eff}$ = 10550 K,
log $g$ = 4.0, [M/H] = 0.0, and $v_{\rm {turb}}$ = 2.0 km/s \citep{soo93,cas03}.\footnotemark[2]

\footnotetext[2]{Adopted from Fiorella Castelli's web page, http://wwwuser.oat.ts.astro.it/castelli/colors.html.
}

Figure 2 compares our isochrones and zero-age HB (ZAHB) models transformed by using the two different
color tables. They are for the chemical compositions and ages appropriate for the most metal-poor
and the most metal-rich (also helium-rich) subpopulations in $\omega$~Cen (see \S3.1 below).
While the models are in general agreement, there are minor but noticeable differences
in the slope of the metal-rich RGB, especially in panel (d), i.e., $B-R$ color.
Assuming that these differences in the Johnson-Cousins $UBVRI$ system would be
similar in the adjacent passbands of the $HST$ filter system, we have applied small corrections
to the synthetic magnitudes of the $HST$ ACS/WFC F435W and F625W passbands.
For instance, for the F435W band, we added the differences in the $B$ band magnitudes between the two tables,
to the synthetic F435W magnitudes. Similarly, we used the differences in the $R$ bands for the
F625W bands. The corrected colors are compared with synthetic colors in panel (e). In case of
the UV (F225W, F275W, and F336W) and the broad VI passbands (F606W and F814W) of the $HST$ WFC3/UVIS
and ACS/WFC systems, the differences between the two color transformations are relatively small (see Fig. 2a and 2c),
and therefore, we did not apply the corrections to these passbands.
The Str\"{o}mgren $b$ and $y$ bands are very similar to the Johnson $B$ and $V$ bands,
and because the two color transformations agree well for the metal-poor
population in $B-V$ color (see Fig. 2b), we also did not apply the corrections to these passbands.
Interstellar extinctions for the filter systems used were estimated according to
\citet{car89} and \citet{sir05}, based on the $E(B-V)$ values listed in the updated catalog of \citet{har96}.

\subsection{A Fitting Technique for Finding Best Parameters}

In order to fit our synthetic models to the observed CMDs and find the values of
ages, metallicities, and helium abundances of subpopulations in each GC,
we have combined both eye fitting on the CMDs and $\chi^2$ minimization technique on the Hess diagrams.
First, from the literature, we predetermined population ratio of subpopulations in a GC,
metallicity ([Fe/H]) of each subpopulation, and a difference in the total CNO content
($\Delta$[CNO/Fe]) between subpopulations. Synthetic CMDs were then constructed
for the relevant filter systems, and they were first compared with the observed CMDs by eye fitting.
At this stage, initial guesses of age and helium abundance were assigned, and distance modulus,
reddening, and [Fe/H] values of each subpopulation were fixed within the observational uncertainty.

The best fitting values of age and helium abundance for all subpopulations were then simultaneously
determined by performing a $\chi^2$ minimization between the observed and synthetic Hess diagrams.
This was inspired by the techniques widely used in dwarf galaxy studies \citep[e.g.,][]{har01,har04,
tol09,rub11}. For example, Figures 3 and 4 illustrate our procedure for NGC~1851.
Panel (a) in Figure 3 shows the CMD regions used in the $\chi^2$ minimization,
and panels (b) and (d) display the Hess diagrams of the observed CMD.
Using the predetermined parameters of each GC, we constructed a set of synthetic CMDs with
an age resolution of 0.1 Gyr and helium abundance resolution of 0.01, including observational error
simulations.\footnotemark[3] Based on these models, synthetic Hess diagrams were produced by averaging
1000 simulations to minimize stochastic effects.
The synthetic Hess diagrams were finally compared with the observed ones using the IDL\footnotemark[4]
function XSQ\rule{0.25cm}{0.4pt}TEST, which performs the $\chi^2$ goodness-of-fit test
between the observed and expected frequencies of a theoretical distribution.
We have fixed helium abundance of the metal-poor first-generation (G1) subpopulation to be normal value
(i.e., Y = Y$_p$ + 2Z, Y$_p$ = 0.23), so that G1 was examined only for age variation,
while the metal-rich second-generation (G2) subpopulation was investigated for the variations of
both age and helium abundance, simultaneously. Panels (c) and (e) in Figure 3 are the model Hess diagrams
with the minimum $\chi^2$, and Figure 4 is the distribution of reduced $\chi^2$ ($\chi_{\nu}^2$)
values for the variations in age and helium abundance of each subpopulation.
The 68\% (1 $\sigma$) confidence levels were calculated by bootstrapping with 1000 simulations
for each subpopulation. The estimated 1 $\sigma$ errors are typically $\pm$ 0.3 Gyr for the relative age
and $\pm$ 0.02 for helium abundance (Y). Note that these errors are only for the case
when the metallicities and the total CNO contents are held fixed, and therefore,
the actual errors would be larger if the measurement errors of metallicity and [CNO/Fe] are not negligible.
The uncertainties in the distance modulus and reddening, which are fixed in the analysis,
as well as those in stellar models, are also not included in the quoted errors.
Note, however, that they would have only limited impact on the estimation of relative differences
in age and helium abundance.

\footnotetext[3]{We adopted the Salpeter initial mass function, with the standard index ($x$ = 1.35),
in our construction of synthetic CMDs. The choice of $x$ has only a negligible effect in our analysis,
since we are dealing with the relatively bright parts of the CMD where the mass range is small in old
stellar populations.
}

\footnotetext[4]{IDL is the Interactive Data Language, a product of ITT Visual Information Solutions: http://www.ittvis.com.
}

\section{SYNTHETIC COLOR MAGNITUDE DIAGRAMS}

\subsection{$\omega$~Cen}

As numerous photometric and spectroscopic studies have been devoted to it, the most massive GC, $\omega$~Cen,
is indeed the most complex object among GCs in the Milky Way. It shows several prominent features in the CMDs
such as a double MS, multiple subgiant branches (SGBs), multiple RGBs, and extreme blue HBs (EBHBs),
as well as large variations in the spectroscopic abundance of heavy elements \citep{lee99,pan00,rey04,fer04,
bed04,sol05a,sol06,sol07,pio05,vil07,bel10,fre75,nor96,sun96,joh10,car10a,mar11a}.
The multiple and discrete RGBs, together with their spectroscopic variations in the abundance of heavy elements,
directly indicate that subpopulations in $\omega$~Cen were influenced by SNe.
Meanwhile, the presence of a double MS and the fact that the bluer MS is more metal-rich
than the redder MS strongly suggest that there is a large variation of helium abundance in this cluster
\citep{bed04,nor04,lee05,pio05}. The presence of unusually extended HB is also naturally explained by the subpopulations
enhanced in helium \citep{lee05}.

Following the techniques and spirits employed in our previous investigation \citep{lee05},
we have constructed new synthetic CMDs for $\omega$~Cen. As described in \S2.1, there have been
two major improvements in our modeling. One is the adoption of the new Y$^2$ HB evolutionary tracks,
together with the corresponding Y$^2$ isochrones, with which the enhancements in both helium abundance
and [CNO/Fe] can be tested self-consistently. The other is the improvement of the color-temperature
transformation including the ``semi-empirical corrections" for the $HST$ ACS/WFC F435W and F625W passbands
(see Fig. 2e). Furthermore, the best fitting parameters are now determined with the method described in \S2.2.
In the construction of synthetic CMDs, the metallicities ([Fe/H]) for the five subpopulations were first adopted
from the most recent spectroscopic observations of RGB stars \citep{joh10,mar11a}, which are largely
consistent with earlier estimates \citep{nor96,sun96,smi00,pan02,rey04,sol05a,sol05b,pio05,vil07,joh08,joh09}.
The differences in [CNO/Fe] among the five subpopulations were taken from the recent observations by
\citet{mar12a}, and the adopted values are listed in table 2.
We have then determined the best fitting ages and helium abundances by performing the $\chi^2$
minimization on the Hess diagrams employing the $HST$ ACS/WFC F435W and F658N passbands.
In Figures 5 and 6, our synthetic models are compared with the observed CMDs by \citet{bel10}, and
the best fitting parameters obtained by our model simulations are listed in Table 2.
These model CMDs were specifically constructed to reproduce the observed CMDs,
adopting the same $HST$ filter systems, population ratios, and photometric errors.

For the two most metal-poor subpopulations, which we refer to as G1 and G2, reasonable agreements
with the observations, from MS to HB, are achieved with normal helium abundances
(i.e., Y = Y$_p$ + 2Z, Y$_p$ = 0.23). The RR Lyrae stars belonging to these metal-poor subpopulations
are also providing a very firm constraint ($\Delta$Y $\lesssim$ 0.01) on the helium abundance
as the luminosities and periods of RR Lyraes depend very sensitively on helium abundance (see below).
For the ages of these subpopulations, our fitting technique is suggesting 13.1 $\pm$ 0.2 and
13.0 $\pm$ 0.3 Gyrs, respectively, and therefore they could be coeval to within the uncertainty.
For the intermediate metallicity subpopulation, G3, however,
a large amount of helium enhancement ($\Delta$Y = 0.18 $\pm$ 0.02)
is suggested from our fitting technique to reproduce not only the bluer and fainter MS but also the EBHB stars.
This is readily understood by the effects of helium abundance in stellar astrophysics.
Because of the higher core temperature and decreased envelope opacity, helium-rich stars, in general,
are bluer and brighter for a given mass. Since they evolve faster than helium-poor stars, helium-rich stars
would have smaller masses for a given age, and therefore the helium-rich MS appears both bluer and fainter
than the helium-poor sequence on the isochrone \citep[see, e.g.,][]{cox68,ibe68,lee05}.
For the same reason, helium-rich HB stars have smaller total masses than do less helium-rich stars
and end up on the blue HB or EBHB \citep{lee94}. The best fitting age for this helium enhanced G3 is
12.0 $\pm$ 0.4 Gyr, which is about 1 Gyr younger than the reference subpopulations (G1 and G2).
Similarly, for the two most metal-rich subpopulations, G4 and G5,
our models require helium enhancements as much as that for G3, from the presence of EBHB stars
and the round shape of the SGBs. If they were not enhanced in helium abundance,
they should have red HB stars or red clumps and more flattened SGBs
at their metallicities, [Fe/H] $\gtrsim$ $-$1.0 (see Fig. 8 below, \citealp[see also][]{sol05b}).
For these subpopulations, our fitting technique is suggesting a common age of 11.4 $\pm$ 0.4 Gyr,
and therefore, the age difference between G1 and G5 is predicted to be about 1.7 $\pm$ 0.5 Gyr.
This formation time scale of $\omega$~Cen has important implications for understanding the behaviors of
the $\alpha$-elements and $s$-process elements \citep[e.g.,][]{nor95,bus99,smi00,joh10,dan11,mar11a,mar12a,
dor11,val11}. Note that the variation in total CNO content between the subpopulations has the most
significant impact on this age difference. As shown in Figure 7, if there were no variation in [CNO/Fe],
our models suggest that all of the subpopulations in $\omega$~Cen would be almost coeval
\citep[see also][]{maf10,dan11,mar12a}.

In order to illustrate, independently, the effects of metallicity, [CNO/Fe], age, and helium abundance
on the overall morphology in CMD, Figure 8 compares the synthetic CMDs constructed under different
combinations of these parameters. The presented models are for the metallicities comparable
to the most metal-poor and the most metal-rich subpopulations in $\omega$~Cen, G1 and G5.
First, the $\Delta$Z-only model in panel (a) is for the case in which G5 has enhanced metallicity
but the same [CNO/Fe], age, and helium abundance with G1. The $\Delta$Z + $\Delta$CNO + $\Delta$t model
in panel (b) is for the case in which [CNO/Fe] is also enhanced and age is decreased for G5.
Both enhanced [CNO/Fe] and decreased age are pushing the HB of G5 to even redder color \citep[see][]{lee94},
while they have opposite effects on the MS turn-off (MSTO) and SGB regions (see also Fig. 7) so that
no appreciable change is seen there. Finally, $\Delta$Z + $\Delta$CNO + $\Delta$t + $\Delta$Y model in panel (c)
is for the case where the helium abundance of G5 is also significantly enhanced, as in our best simulation.
As is clear from Figure 8, G5 would have the red HB in the models without helium enhancement,
while the enhanced helium abundance in panel (c) pushes the HB stars into the EBHB region.
It is also shown that the luminosity and color of MSTO, together with the SGB slope, are
fairly affected by this variation in helium abundance.

In our modeling, the effects of enhancements in helium and the total CNO abundances were only considered
in the stellar interior, and it was assumed that their effects on stellar atmosphere would be negligible.
For the effect of helium abundance, we have confirmed this using the helium-enhanced model fluxes
of \citet{cas03}.\footnotemark[6] Figure 9a shows the model spectra of two stars with
different helium abundances ($\Delta Y$ = 0.10) but the same temperature, gravity, and metallicity
(i.e., $T \rm_{eff}$ = 6000 K, log $g$ = 4.5, and [Fe/H] = $-$1.5),
which correspond roughly to stars near the MSTOs of G2 and G3 in $\omega$~Cen.
We can see that a large variation in helium abundance ($\Delta Y$ = 0.10) has only negligible effect
on stellar atmosphere \citep[see also][]{gir07,sbo11}. In the UV region, for the wavelengths shorter than $\sim$4000 \AA,
however, some differences between the two synthetic spectra are noticed. Figure 9c shows that,
in the worst situation, combinations of $T \rm_{eff}$, log $g$, [Fe/H], and the helium variations of
$\Delta Y$ = 0.10 can make differences in magnitude up to $\sim$0.15 in the UV bands.
In the case of the total CNO abundance, our assumption appears to be valid also in the optical passbands.
Using synthetic spectra, \citet{sbo11} have shown that the variations in C, N, O, and Na abundances
mainly affect wavelengths shorter than $\sim$4000 \AA. They have also reported that the effect on the
$hk$ index is negligible (at most $\sim$0.04 mag), while CMDs including $U$ band could be affected by
$\sim$0.2--0.3 mag for the $\sim$0.3 dex variation in [CNO/Fe]. Considering these uncertainties
in the UV regime, the best fitting parameters in our models are determined based only
on the optical bands and the $hk$ index.

\footnotetext[6]{The model atmospheres for enhanced helium abundances can be downloaded at
Fiorella Castelli's web page, http://wwwuser.oat.ts.astro.it/castelli/grids.html.
They present the synthetic spectra for three metallicity grids, [Fe/H] = $-$1.5, $-$0.5, and 0.5,
and two gravity grids, log $g$ = 1.5 and 4.5, for $\Delta$Y = 0.10.
For $\Delta$Y = 0.20, only metal-rich ([Fe/H] = 0.5) models are available.
}

Some HB stars in our models are in the RR Lyrae instability strip, and they can provide additional
constraints in the modeling. Following \citet{lee90}, we have computed the fundamental periods (P$\it{_f}$),
using the equation of stellar pulsation \citep{van71}, for the RR Lyrae candidates in the synthetic HB.
Figure 10 compares our synthetic stars to the observed RR Lyrae variables in $\omega$~Cen,
where the metallicities of 74 RR Lyrae stars are from the spectroscopic measurements by \citet{sol06}.
In panel (a), the periods of $c$-type RR Lyrae stars are fundamentalized assuming the period ratio,
P$_c$/P$_{ab}$ = 0.745, between the $c$-type and $ab$-type RR Lyare stars \citep{cle01,alc00}. In our models, only
the two most metal-poor subpopulations, G1 and G2, have RR Lyrae stars. The HB stars from the three
metal-rich and super-helium-rich subpopulations, G3, G4, and G5, are too hot to produce RR Lyare stars
as they are on the EBHB region far beyond the blue edge of the instability strip.
In panel (b), metallicities of the synthetic RR Lyrae stars have been assigned randomly
according to a Gaussian distribution centered on the metallicities of the G1 and G2 (see Table 2), respectively,
with $\sigma_{[Fe/H]}$ = 0.13, which is roughly consistent with the observational error in [Fe/H] measurement.
The predicted periods, $\it V$ magnitudes, and [Fe/H] from our model populations, G1 and G2,
show good agreements with the observed RR Lyrae variables except for the several stars having intermediate metallicity
($-$1.4 $<$ [Fe/H] $<$ $-$1.0). As pointed out by \citet{sol06}, these stars are expected to have normal helium abundance
because their luminosities and periods are not increased \citep[see][]{lee94}. This implies that a minor subpopulation
having a similar metallicity but different helium abundance with G3 may coexist in $\omega$~Cen. The presence of these
normal-helium, intermediate metallicity RR Lyrae stars, together with the $HST$ WFC3 UV photometry by \citet{bel10},
suggests that $\omega$~Cen contains additional minor subpopulations between the five subpopulations adopted in this study.

\subsection{M22}

It has long been suspected that RGB stars in the massive GC M22 show a spread in heavy element abundance,
but the presence of differential reddening, caused by its location near the Galactic plane, has prevented
a firm confirmation of its reality \citep[e.g.,][]{hes77,pil82,nor83,leh91,ant95,mon04,iva04}.
Recent observations, however, have shown that M22 indeed hosts two stellar subpopulations
differing in the abundance of heavy elements. Based on high resolution spectroscopy, \citet{mar09,mar11b}
identified two stellar groups separated by $\sim$0.15 dex in [Fe/H],
while \citet{dac09} reported a somewhat larger difference in metallicity ($\sim$0.26 dex).
Meanwhile, \citet{jwlee09a} has clearly shown that M22 has a discrete double RGB having different Ca abundance from
the $hk$ index of the {\it Ca-by} photometry.

In the left panels of Figure 11, we show CMDs of M22 from S.-I. Han et al. (2012, in preparation),
which were observed with the Str\"{o}mgren $b$, $y$ and a new narrow band calcium filter
specially designed to avoid a possible contamination from the adjacent CN bands.
We can see two well-separated RGBs in the $hk$ index, which are very similar to those
in Figure 1 of \citet{jwlee09a}, and a bimodal distribution of HB stars composed of blue HB and EBHB.
Based on these photometric data and the CMD from the ACS survey \citep{pio09}\footnotemark[5], shown in Figure 12a,
we constructed synthetic CMDs as described in \S2.
The difference in [CNO/Fe] between the two subpopulations was adopted to be 0.13 dex from the recent observations
\citep[][see also \citealp{alv12}]{mar11b,mar12b}.

\footnotetext[5]{The photometric data are available at the web page of the ACS survey of Galactic GCs \citep{sar07,and08},
http://www.astro.ufl.edu/\~{}ata/public\_{}hstgc/databases.html.
}

The right panels of Figures 11 and 12 are the final outcomes of our population models compared with
the observed CMDs. The best fitting model parameters are listed in Table 3.
The metal-poor subpopulation, which we refer to as the first-generation (G1) population,
contains the bluer RGB, brighter SGB, and blue HB stars, while the metal-rich subpopulation,
which we refer to as the second-generation (G2) population, has the redder RGB, fainter SGB,
and EBHB stars. In order to simultaneously reproduce these observed splits on the RGB and SGB, and
the bimodal HB distribution, we need a small but important difference in metallicity
($\Delta$[Fe/H] = 0.25 dex), and at the same time, a large difference in helium abundance
($\Delta$Y = 0.09 $\pm$ 0.04) between the two subpopulations. By applying a small difference in
the total CNO content ($\Delta$[CNO/Fe] = 0.13 dex; \citealt{mar12b}), our models suggest that the best fitting
age difference is very small ($\Delta$t = 0.3 $\pm$ 0.4 Gyr).
The adopted difference in metallicity is similar to the observed difference obtained from Ca II triplet
($\sim$0.26 dex; \citealt{dac09}), while it is somewhat larger than the difference obtained from
[Fe/H] ($\sim$0.15 dex; \citealt{mar09,mar11b}). This is probably due to a small increase in calcium
abundance ($\Delta$[Ca/Fe] $\approx$ 0.1; \citealt{mar09,mar11b}) in the metal-rich subpopulation, G2.
Our models are also consistent with the recent spectroscopic measurements by \citet{mar12b},
which have shown that stars in the faint SGB are more metal-rich compared to the stars in the bright SGB.

Figure 13 explains the effects of variations in the total CNO abundance, overall metallicity, and
helium abundance, independently. If we assume only the difference in [CNO/Fe]
between the two subpopulations, to reproduce the split on the SGB,
as suggested for NGC~1851 \citep{cas08,sal08}, our models fail
to explain the presence of the RGB split in the $hk$ index and the EBHB stars (the upper panels).
Note that, for illustrative purpose, we have increased the difference to 0.5 dex
from the observed value ($\sim$0.13 dex), because the observed difference is too small to
reproduce the SGB split. Similarly, with the assumption of enhancement only in [Fe/H],
our model for G2 also fails to reproduce the EBHB stars, while the split on the RGB can be explained
(the middle panels). This is a natural consequence of the metallicity effect, as the HB morphology is
getting redder with increasing metallicity \citep[see, e.g.,][]{lee94}. While this difference in
[Fe/H] reproduces some split on the SGB, it is not sufficient to explain the observed difference.
The bottom panels of Figure 13 show that additional enhancement in helium is needed to overcome
the effect of metallicity on the HB and to reproduce the EBHB stars, as already explained in \S3.1.
Similarly to the case of $\omega$~Cen, this enhancement of helium also helps
to reproduce the observed split on the SGB (see Fig. 12d).

The argument based on the population ratio further supports our hypothesis that the EBHB stars belong to the metal-rich
subpopulation G2. The number ratio between the bluer and redder RGBs in Figure 11c and those in \citet{jwlee09a}
is about 0.7:0.3. \citet{pio09} and \citet{mar09} have also suggested a similar number ratio
(0.62:0.38) between the bright and faint SGBs. Within the error, this is also consistent with
the ratio of the EBHB stars which comprise roughly 25\% of all HB stars in panel (a) of Figures 11 and 12.
This would then suggest that the bright SGB, bluer RGB, and blue HB stars are all associated with
the majority, metal-poor subpopulation (G1), while the faint SGB and redder RGB stars are progenitors
of the EBHB stars, associated with the minority metal-rich subpopulation (G2).

\subsection{NGC~1851}

Recent observations of NGC~1851, which is well-known for its unusual bimodal HB distribution
\citep[e.g.,][]{sav98,wal98}, indicate that it also belongs to the peculiar group of massive GCs
influenced by SNe enrichment like $\omega$~Cen and M22. Not only a double RGB has been revealed
from several photometric observations \citep{jwlee09a,jwlee09b,han09,car11a},
but also some spreads in [Fe/H] have been reported from spectroscopic observations \citep{car10c,car11b}.
NGC~1851 also shows other signatures for the presence of two subpopulations, such as a double SGB \citep{mil08},
a CN bimodality \citep{hes82}, and variations in light and $s$-process element abundances \citep{yon08,yon09,vil10,car11b}.
The origin of the observed SGB split, together with its possible links to the RGB split
and the bimodal HB distribution, however, is still under debate. For instance, the SGB spilt can be explained
by a difference in age of $\sim$1--1.5 Gyr \citep{mil08,car10c,car11b,gra12b}, by an increase of total CNO abundance
by a factor of $\sim$2--4 \citep{cas08,sal08,yon09,ven09,dan09}, or by a combined effect of the differences
in overall metallicity ($\Delta$[Fe/H] $\approx$ 0.15 dex) and helium abundance ($\Delta$Y $\approx$ 0.05; \citealp{han09}).

Based on the framework of our previous modeling \citep{han09}, we have constructed new synthetic CMDs
for NGC~1851. The best fitting parameters were determined with the technique described in \S2.2.
In the case of NGC~1851, there is still no general consensus on the observed value of
$\Delta$[CNO/Fe] between the two subpopulations \citep{yon09,vil10,lar12}, but \citet{gra12b} recently
concluded that the difference in [CNO/Fe] is less than 0.2 dex. Considering this and the similarity
of NGC~1851 with M22, here, we have adopted 0.1 dex for $\Delta$[CNO/Fe]. Given the small value,
this should have only a little effect in our modeling.

In Figure 14, our models are compared with new CMDs of NGC~1851, obtained from {\it Ca-by}
photometry (S.-I. Han et al. 2012, in preparation). This observation confirms the previously reported
split of the RGB in the $hk$ index \citep{jwlee09a,jwlee09b}. Similarly, Figure 15 compares our models with
the CMDs from the ACS survey \citep{mil08}.\footnotemark[5] The model parameters determined by our fitting
technique are listed in Table 4. With the small variations in metallicity ($\Delta$[Fe/H] = 0.13 dex) and
the total CNO content ($\Delta$[CNO/Fe] = 0.1 dex) between the two subpopulations, our fitting technique
suggests that the difference in helium abundance is considerable ($\Delta$Y = 0.06 $\pm$ 0.01),
while the difference in age is relatively small ($\Delta$t = 0.5 $\pm$ 0.2 Gyr).
This result is qualitatively similar to the case of M22.
The faint SGB can be explained mostly by the combined effects of the metallicity and helium enhancements,
as both of them make MSTOs fainter (Fig. 15), while the small variations in the total CNO abundance and age
have only little impact on the CMD. In particular, as discussed above (Fig. 7), the total CNO content
and age basically affect MSTOs in opposite directions. If there were no difference in [CNO/Fe]
between G1 and G2, then the age difference would be even smaller ($\Delta$t $\approx$ 0.0 Gyr).
The Ca-strong redder RGB in the $hk$ index can be accounted for solely by the effect of metal
enhancement, which makes RGBs redder (bottom panels of Fig. 14). The helium enhancement slightly moves
the RGB to the opposite direction, but the effect of metal enrichment is far more important.
In the HB region, the effect of helium enhancement, which makes HB morphology bluer,
easily overcomes the effects of metallicity and CNO contents (see Fig. 13), and therefore
the presence of blue HB, populated by G2, can be explained. Note that our model
for G2 is also consistent with recent measurement of helium abundances for the blue HB stars
(Y = 0.29 $\pm$ 0.05; \citealp{gra12b}).

Our models indicate, therefore, that the majority subpopulation, G1, comprises the metal-poor bright SGB,
Ca-weak bluer RGB, and red HB stars; while the minority subpopulation, G2, is associated with the
metal-rich faint SGB, Ca-strong redder RGB, and blue HB stars.
This is consistent with the population ratios estimated from Figures 14 and 15.
The number fraction between the bright and faint SGBs is roughly 0.68:0.32 (Fig. 15c), and
the ratio between the bluer and redder RGBs is about 0.71:0.29 (Fig. 14c).
This also agrees with the number ratios reported in previous works \citep{mil08,mil09,jwlee09a,han09},
to within the uncertainty. The ratio between the red and blue HBs is about 0.69:0.31 (Fig. 14a and Fig. 15a),
if RR Lyrae stars belong to metal-poor subpopulation (G1). Note that the mean period of RR Lyrae stars
($<$P$_f$$>$ = 0.531 day), which is normal for the metallicity of NGC~1851 \citep{wal98,cle01}, also indicates
that most of the RR Lyrae variables come from the majority subpopulation (G1) having normal helium abundance.


\section{DISCUSSION}

We have compared our models with the observations of three peculiar GCs, $\omega$~Cen, M22, and NGC~1851,
to estimate ages, metallicities, and helium abundances of their subpopulations.
Based on our results, in Figure 16, we schematically illustrate the star formation histories and accompanying
enrichments of metallicity and helium abundance in these stellar systems. We find that the age differences
between the metal-rich and metal-poor subpopulations in these GCs are relatively small, $\sim$0.3--1.7 Gyr,
and metal-rich subpopulations with redder RGBs are also enhanced in helium abundance.
The formation time scale of stellar populations in these GCs is therefore expected to be fairly short,
i.e., $\sim$1.7 Gyr for $\omega$~Cen and less than 1 Gyr for M22 and NGC~1851.

How this could be understood in terms of their formation origin?
The most natural assumption would be that they are self-enriched, in which later generations of stars formed
from the gas enriched by the ejecta of massive stars in earlier generations.
As for the sources of chemical enrichment and pollution in these GCs, at least three viable mechanisms have been
proposed thus far. The explosion of type II SNe is required for the enrichments of heavy elements \citep{tim95}
and perhaps helium \citep{nor04,pio05}, while winds from the FRMSs and IMAGBs are suggested for the pollution
of light elements and the enhancement of helium \citep{dec07,dan07,ven08}.
The enhancement of the total CNO abundance in the metal-rich later generation subpopulations would also
indicate the contribution by type II SNe \citep[see also][]{mar12a}.
Considering the order of time required to emerge for these mechanisms, a possible scenario would be as follows:
(1) Formation of the first-generation metal-poor (bluer RGB) stars from the gas having normal helium and
light element abundances. (2) Pollution of the remaining gas by the winds from FRMSs, which enhance helium
and alter the abundance profile of light elements. (3) The most massive ($\gtrsim$ 8 M$_\odot$) stars then
would explode as type II SNe, enriching overall metallicity, including heavy elements and the total CNO content.
(4) Further pollution of the gas by the ejecta from IMAGB stars ($\sim$3--7 M$_\odot$) would follow,
which enhance helium and change light element abundances. (5) Finally, formation of the second-generation
metal-rich (redder RGB) stars from the gas now enriched in overall metallicity and helium and polluted in
light element abundances.
While this scenario can naturally explain the presence of metal-rich and helium-enhanced subpopulations
in these GCs, this requires that they were much more massive in the past, because their present masses
are too small to retain the ejecta of SNe explosions \citep[e.g.,][]{dop86,bau08}. This suggests that
these GCs were once nuclei of primeval dwarf galaxies and then merged and dissolved in the proto-Galaxy,
as is widely accepted for $\omega$~Cen \citep[e.g.,][and references therein]{lee99,bek03,pio05,mez05,lee07,
jwlee09a,bau08,fer09,pfl09,coh10,bek12a}. Note that a significant fraction ($\sim$30\%) of the helium-enhanced
subpopulation in these GCs is also best explained if the second generation stars were formed from enriched gas
trapped in the deep gravitational potential well of the ancient dwarf galaxies \citep{bek06}.

Recently, \citet{car10c,car11b} and \citet{bek12b} suggested that NGC~1851 might have formed by simple merging of
two GCs having different heavy element abundances initially belonged to a proto-dwarf galaxy. While this scenario
can explain the presence of RGB split and the spread in heavy element abundance without assuming self-enrichment,
this is not easily reconciled with one of our main results, which suggests that the metal-rich subpopulation is
also enhanced in helium abundance. In this scenario, it would be difficult to understand why all stars in initially
more metal-rich GCs were selectively enhanced in helium abundance, while those in metal-poor GCs were not.
Note, however, that this merging scenario might be able to explain the absence of helium enhancement between G1 and
G2 in $\omega$~Cen (see \S3.1), where G1 and G2 formed by simple merging of two GCs in the central part of a
proto-dwarf galaxy, while more metal-rich and helium-enhanced subpopulations, G3, G4, and G5, formed later
from the enriched gas trapped in the core of this merged stellar system in a proto-dwarf galaxy.
Despite many uncertainties, all of the above scenarios are possible only in the primeval dwarf galaxy
environment, and therefore, dwarf galaxy origin of these GCs appears to be inevitable.


\acknowledgments We are indebted to the anonymous referee for encouraging us to include the CNO enhanced
models and to employ parameter estimation technique, which led to several key improvements in the manuscript.
We also thank Sang-Il Han for providing the observational data in advance of publication,
Chongsam Na for his assistance in the construction of nitrogen enhanced stellar evolutionary tracks,
and Suk-Jin Yoon for helpful discussions. Support for this work was provided by the National Research
Foundation of Korea to the Center for Galaxy Evolution Research.

\clearpage

\begin{deluxetable}{ll}

\tabletypesize{\scriptsize}
\tablecaption{Model Ingredients\label{tbl-1}}
\tablehead{
\colhead{Ingredient} & \colhead{References}
}
\startdata
Isochrones from MS to RGB            \dotfill & Yonsei--Yale (Y$^2$) isochrones (\citealt{yi08}; Y.-W. Lee et al. 2012, in preparation) \\
HB evolutionary tracks               \dotfill & Yonsei--Yale (Y$^2$) HB tracks (Y.-W. Lee et al. 2012, in preparation)   \\
Color--$T \rm_{eff}$ transformation  \dotfill & \citet{gre87} color table                                \\
                                              & \citet{cas03} model atmosphere                           \\
\enddata
\end{deluxetable}

\clearpage

\begin{deluxetable}{lcccccccc}

\tabletypesize{\scriptsize}
\tablecaption{Parameters suggested from our best simulation of $\omega$~Cen\label{tbl-2}}
\tablewidth{0pt}
\tablehead{
\colhead{Population} & \colhead{$Z$} & \colhead{[Fe/H]\tablenotemark{a}} & \colhead{$\Delta$[CNO/Fe]\tablenotemark{b}}
& \colhead{$Y$} & \colhead{Age(Gyr)} & \colhead{Mass-loss($M_\odot$)\tablenotemark{c}}
& \colhead{Fraction\tablenotemark{d}} & \colhead{other name\tablenotemark{d}}
}
\startdata
G1 \dotfill & 0.0005 & $-$1.81 & 0.0  & 0.231            & 13.1 $\pm$ 0.2 & 0.199 & 0.42 & RGB-MP    \\
G2 \dotfill & 0.0009 & $-$1.55 & 0.0  & 0.232            & 13.0 $\pm$ 0.3 & 0.217 & 0.28 & RGB-MInt1 \\
G3 \dotfill & 0.0015 & $-$1.31 & 0.14 & 0.41 $\pm$ 0.02  & 12.0 $\pm$ 0.4 & 0.171 & 0.17 & RGB-MInt2 \\
G4 \dotfill & 0.0057 & $-$1.01 & 0.47 & 0.38 $\pm$ 0.02  & 11.4 $\pm$ 0.4 & 0.220 & 0.08 & RGB-MInt3 \\
G5 \dotfill & 0.0136  & $-$0.62 & 0.47 & 0.39 $\pm$ 0.02  & 11.4 $\pm$ 0.5 & 0.251 & 0.05 & RGB-a     \\
\enddata
\tablenotetext{a}{[$\alpha$/Fe] = 0.3}
\tablenotetext{b}{From \citet{mar12a}.}
\tablenotetext{c}{Mean mass-loss on the RGB for $\eta$ = 0.53.}
\tablenotetext{d}{From \citet{sol05a} and \citet{pan00}.}
\end{deluxetable}


\clearpage

\begin{deluxetable}{lccccccc}

\tabletypesize{\footnotesize}
\tablecaption{Parameters suggested from our best simulation of M22\label{tbl-3}}
\tablewidth{0pt}
\tablehead{
\colhead{Population} & \colhead{$Z$} &\colhead{[Fe/H]\tablenotemark{a}} & \colhead{$\Delta$[CNO/Fe]\tablenotemark{b}}
& \colhead{$Y$} & \colhead{Age(Gyr)} & \colhead{Mass-loss($M_\odot$)\tablenotemark{c}} & \colhead{Fraction}}
\startdata
G1 \dotfill & 0.00035 & $-$1.96 & 0.0  & 0.231           & 12.8 $\pm$ 0.2 & 0.189 & 0.7 \\
G2 \dotfill & 0.00067 & $-$1.71 & 0.13 & 0.32 $\pm$ 0.04 & 12.5 $\pm$ 0.4 & 0.192 & 0.3 \\
\enddata
\tablenotetext{a}{[$\alpha$/Fe] = 0.3}
\tablenotetext{b}{From \citet{mar12b}.}
\tablenotetext{c}{Mean mass-loss on the RGB for $\eta$ = 0.53.}
\end{deluxetable}

\clearpage

\begin{deluxetable}{lccccccc}

\tabletypesize{\footnotesize}
\tablecaption{Parameters suggested from our best simulation of NGC~1851\label{tbl-4}}
\tablewidth{0pt}
\tablehead{
\colhead{Population} & \colhead{$Z$} & \colhead{[Fe/H]\tablenotemark{a}} & \colhead{$\Delta$[CNO/Fe]\tablenotemark{b}}
& \colhead{$Y$} & \colhead{Age(Gyr)} & \colhead{Mass-loss($M_\odot$)\tablenotemark{c}} & \colhead{Fraction}
}
\startdata
G1 \dotfill & 0.0012 & $-$1.43 & 0.0  & 0.232           & 11.1 $\pm$ 0.1 & 0.208 & 0.7 \\
G2 \dotfill & 0.0018 & $-$1.30 & 0.10 & 0.29 $\pm$ 0.01 & 10.6 $\pm$ 0.2 & 0.212 & 0.3 \\
\enddata
\tablenotetext{a}{[$\alpha$/Fe] = 0.3}
\tablenotetext{b}{From \citet{vil10} and \citet{gra12b}}
\tablenotetext{c}{Mean mass-loss on the RGB for $\eta$ = 0.53.}
\end{deluxetable}

\clearpage
\begin{figure}
\epsscale{0.7}
\plotone{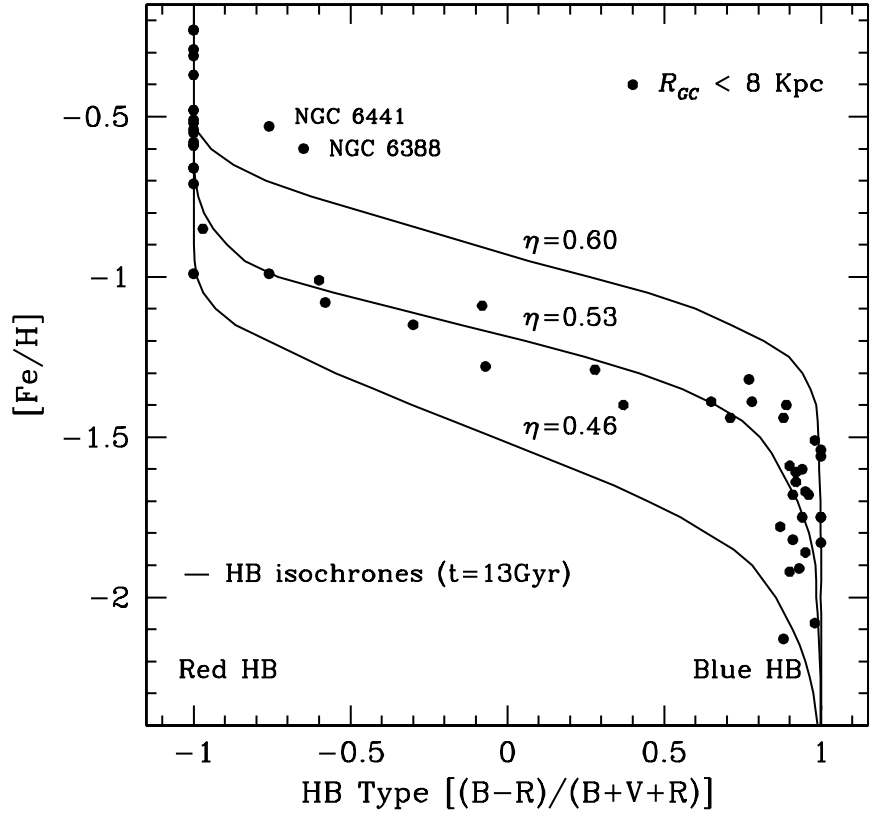}
\caption{
Calibration of the \citet{rei77} mass-loss coefficient, $\eta$, on the HB morphology vs. [Fe/H] relation.
Data for the inner halo (R$_{GC}$ $<$ 8 kpc) GCs from \citet{lee94,lee07}. Solid lines are theoretical isochrones
produced by our synthetic HB models for 13 Gyr using $\eta$ = 0.46, 0.53, and 0.60, respectively.
Two metal-rich GCs with extended HB are marked and are excluded in the comparison.
}
\end{figure}

\clearpage
\begin{figure*}
\epsscale{0.9}
\plotone{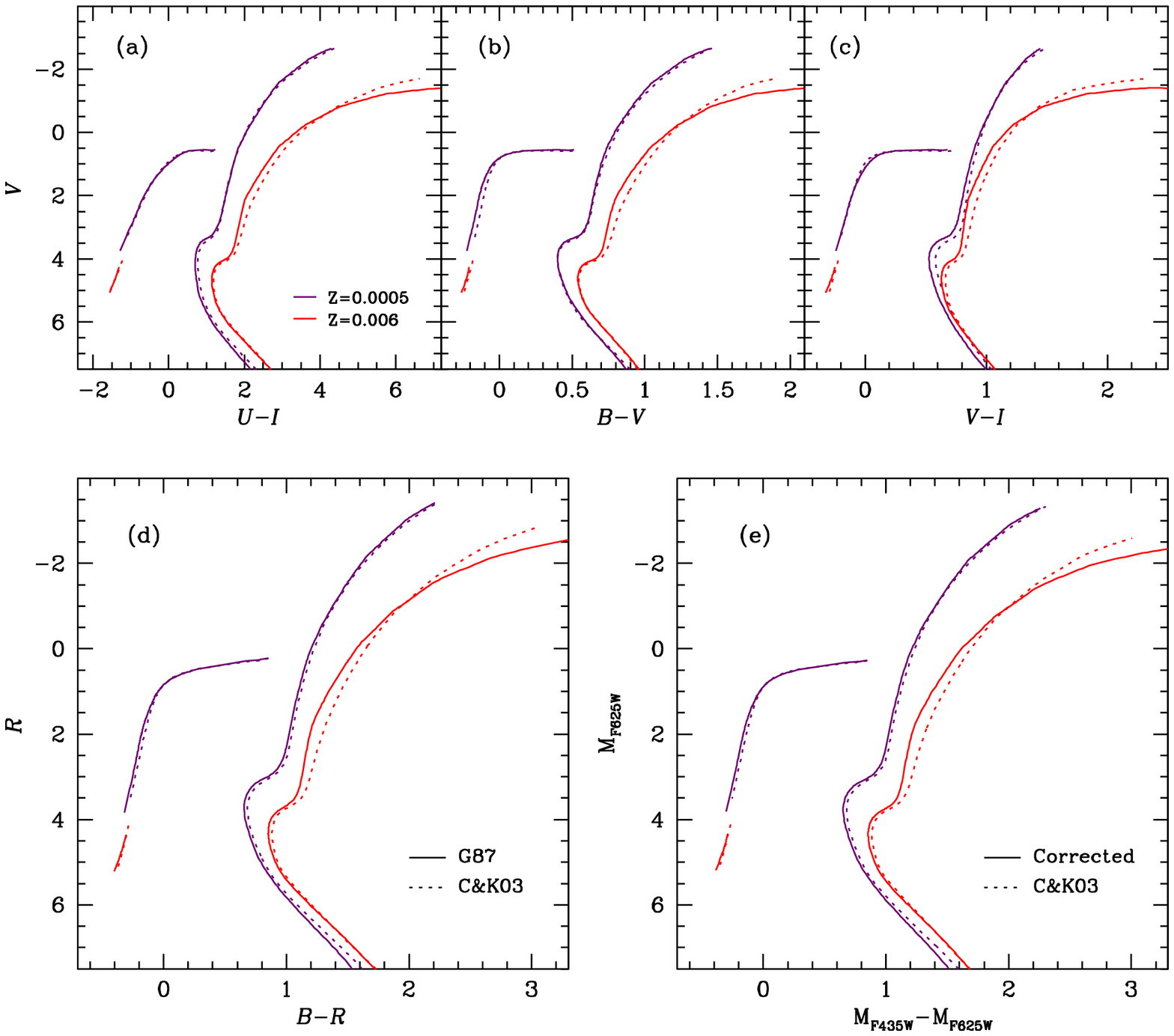}
\caption{
(a)--(d) Comparison of isochrones and ZAHB models constructed with the color--$T \rm_{eff}$ transformations
by \citet[][solid lines]{gre87} and by \citet[][dotted lines]{cas03}.
The model lines plotted are identical to the models for the most metal-poor (purple lines) and the most metal-rich
(red lines) subpopulations in $\omega$~Cen (see \S3.1 below). (e) Solid lines are models corrected by adding the
difference between the two color transformations in (d) to the dotted lines obtained from \citet{cas03}.
}
\end{figure*}

\clearpage
\begin{figure*}
\plotone{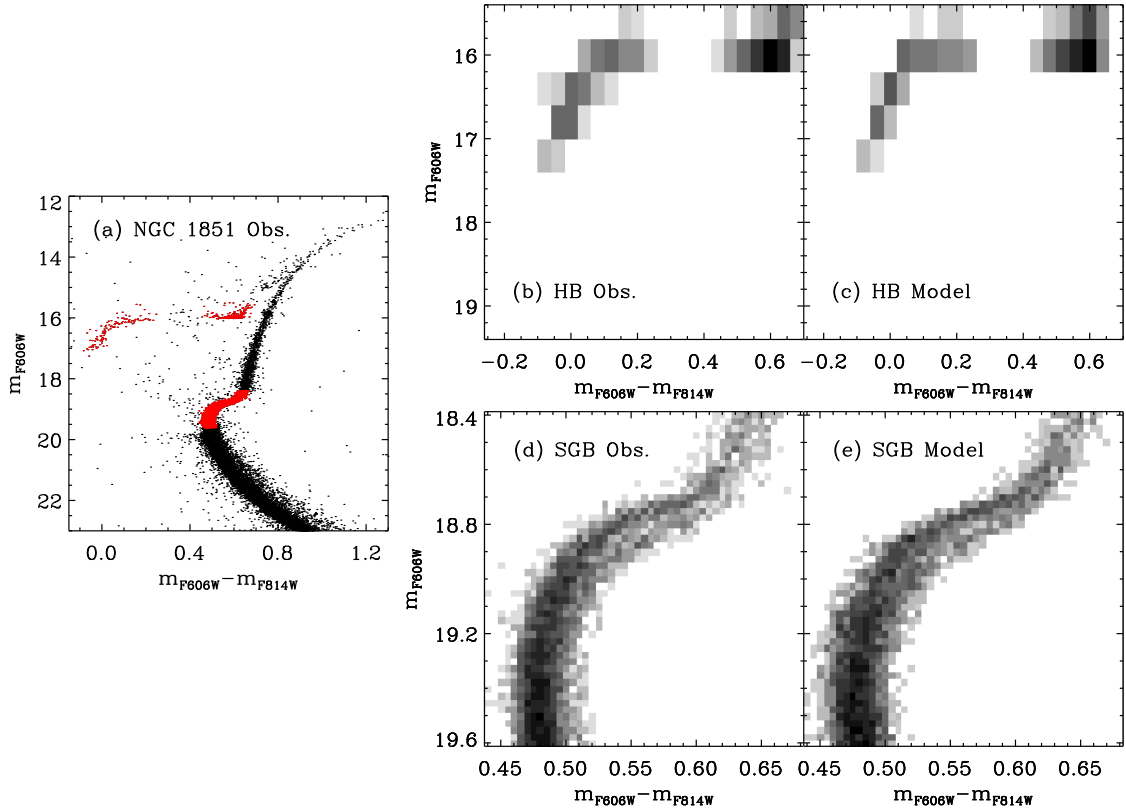}
\caption{
An example of the comparison between the observed and synthetic Hess diagrams used in our $\chi^2$ minimization
technique for NGC~1851. (a) Red points represent CMD regions used in our $\chi^2$ minimization
(data from \citealt{mil08}). (b)--(e) The observed and synthetic Hess diagrams for HB and
SGB regions, respectively.
}
\end{figure*}

\clearpage
\begin{figure*}
\plotone{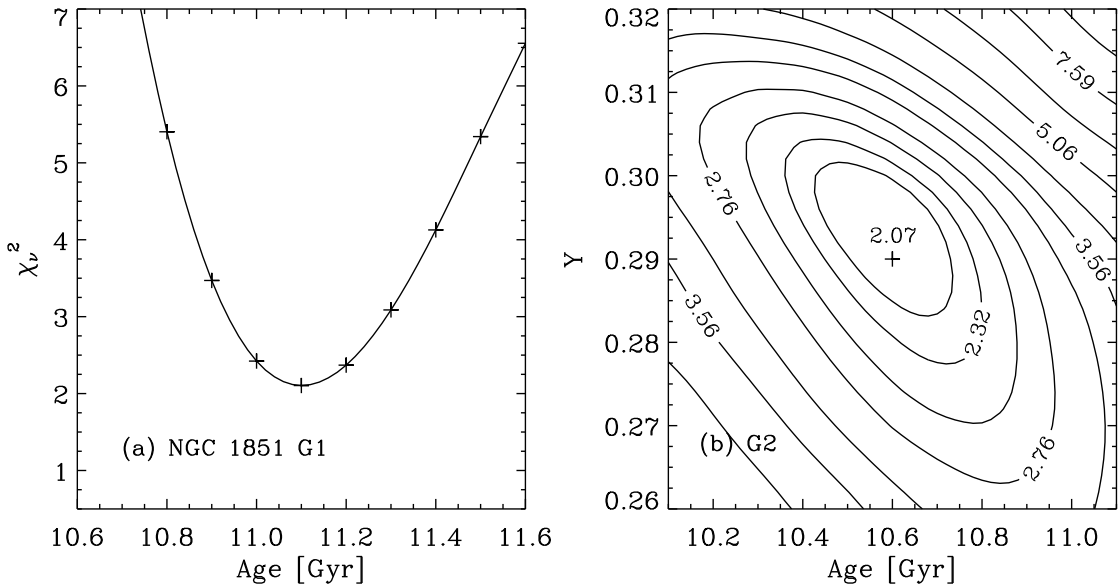}
\caption{
Distributions of reduced $\chi^2$ ($\chi_{\nu}^2$) values in the fitting of our models with
the observed Hess diagram for NGC~1851.
For the first-generation (G1) population, only age is examined,
while for the second-generation (G2) population, both age and helium abundance (Y) are simultaneously
investigated to find the values with minimum $\chi^2$. In panel (b), the values of $\chi_{\nu}^2$
are listed on the contour map (see text).
}
\end{figure*}


\clearpage
\begin{figure*}
\epsscale{0.8}
\plotone{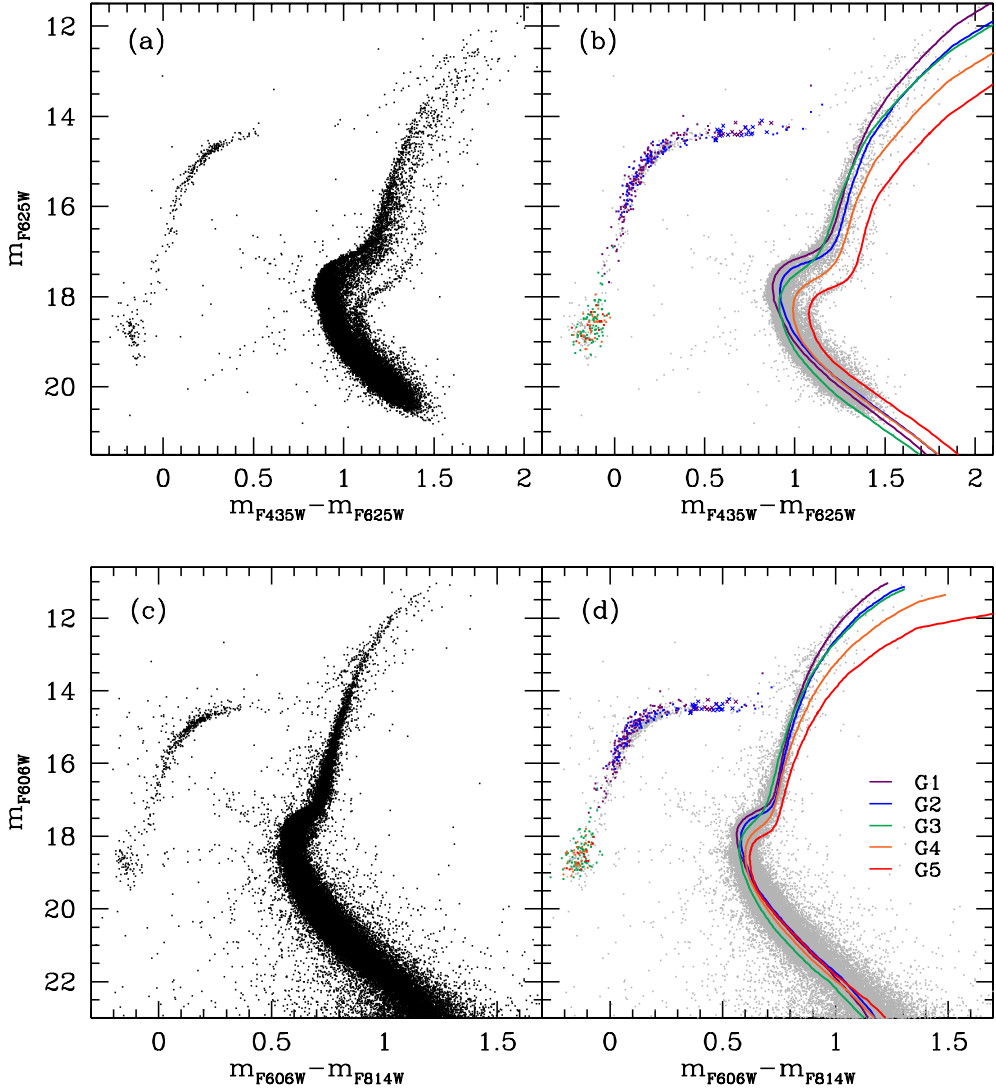}
\caption{
Comparison of our models with the observations for $\omega$~Cen.
(a), (c) Observed CMDs by $HST$ ACS/WFC for the F435W, F625W and F606W, F814W passbands (data from \citealt{bel10}).
(b), (d) Our population models compared on the observed CMDs. Parameters suggested from our best simulation
are listed in Table 4. Adopted distant modulus and reddening are ($m$$-$$M)_{F625W}$ = 14.03, $E(F435W$$-$$F625W$) = 0.225,
and ($m$$-$$M)_{F606W}$ = 14.05, $E(F606W$$-$$F814W)$ = 0.11, respectively.
}
\end{figure*}

\clearpage
\begin{figure*}
\epsscale{0.8}
\plotone{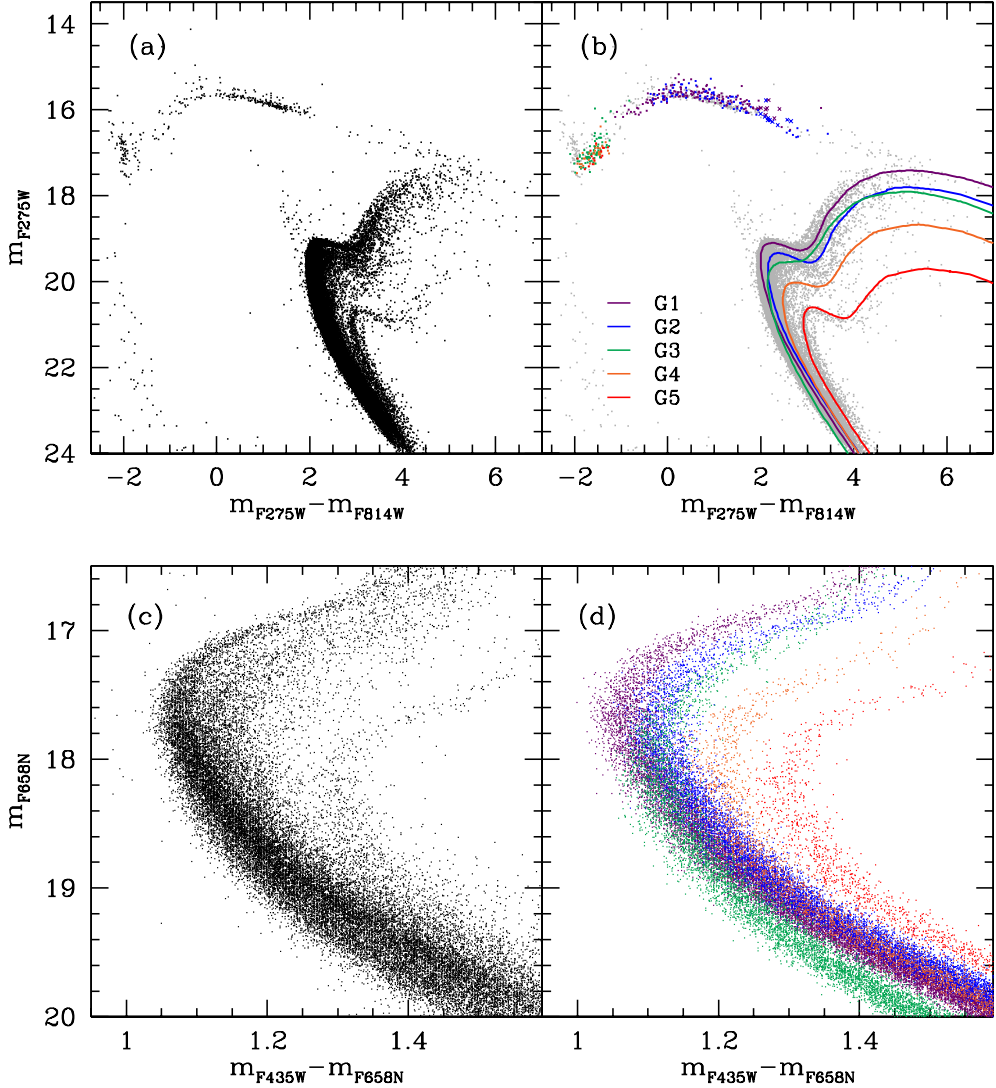}
\caption{
(a), (b) Same as Fig. 5 but for the F275W ($HST$ WFC3/UVIS) and F814W (ACS/WFC) passbands (data from \citealt{bel10}).
We can see that the model parameters obtained from the optical (F435W, F625W) bands can also reproduce the observed
features from the UV (F275W) and broad $I$ (F814W) bands.
(c) CMD for the MS and SGB region in the ACS/WFC F435W and F658N passbands.
(d) Our synthetic models constructed with the photometric errors and total number of stars comparable to the observed CMD
in panel (c). Adopted distant modulus and reddening are ($m$$-$$M)_{F275W}$ = 14.85, $E(F275W$$-$$F814W$) = 0.91,
and ($m$$-$$M)_{F658N}$ = 13.97, $E(F435W$$-$$F658N)$ = 0.22, respectively.
}
\end{figure*}

\clearpage
\begin{figure}
\epsscale{0.4}
\plotone{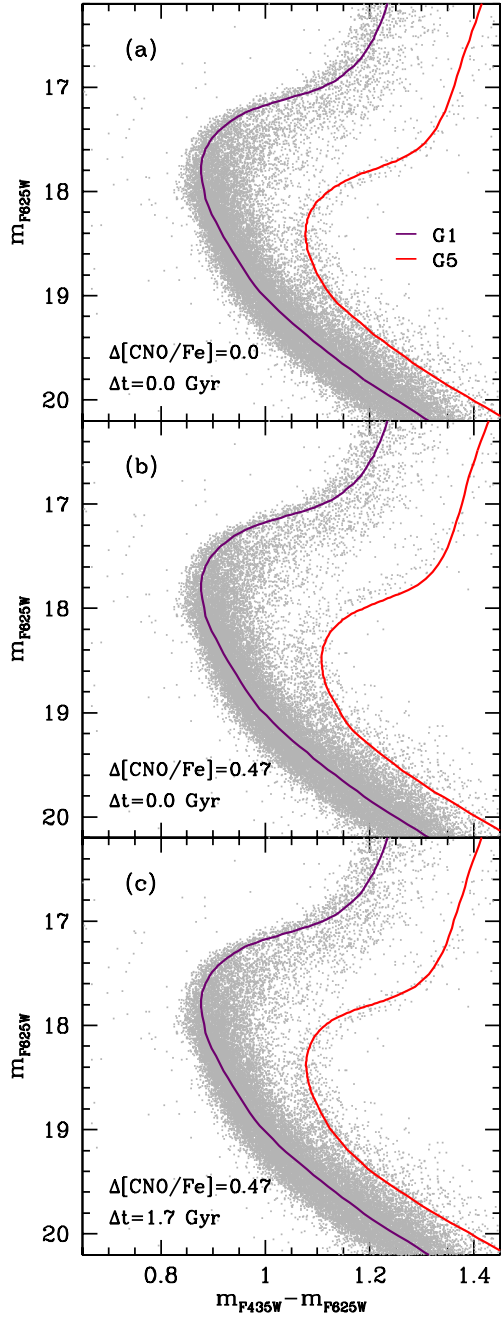}
\caption{
Similar to Fig. 5b, but here the models are constructed to illustrate the effects of $\Delta$[CNO/Fe]
on the age dating. (a) If $\Delta$[CNO/Fe] = 0.0 between G1 and G5, no difference in age is
obtained between them. (b), (c) When the observed $\Delta$[CNO/Fe] is included, $\Delta$t = 1.7 Gyr
is required between G1 and G5, in the sense that G5 is younger, otherwise models would not match
the observations.
}
\end{figure}

\clearpage
\begin{figure}
\epsscale{0.4}
\plotone{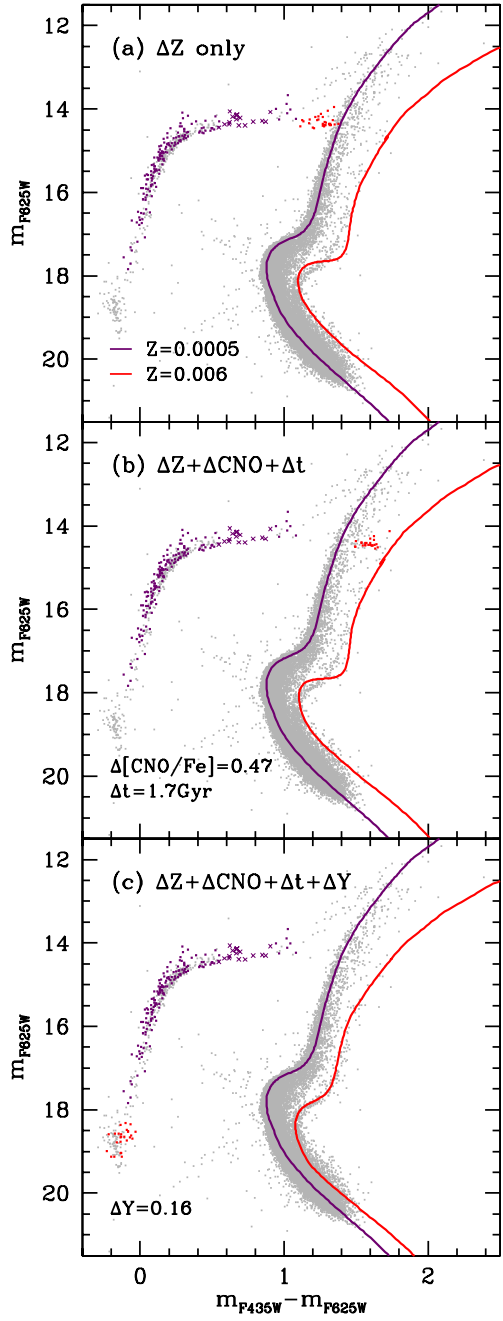}
\caption{
Similar to Fig. 5b, but here the models are constructed to illustrate, independently, the effects of
metallicity, the total CNO content, age, and helium abundance on the overall morphology in the CMD (see text).
}
\end{figure}

\clearpage
\begin{figure*}
\epsscale{0.9}
\plotone{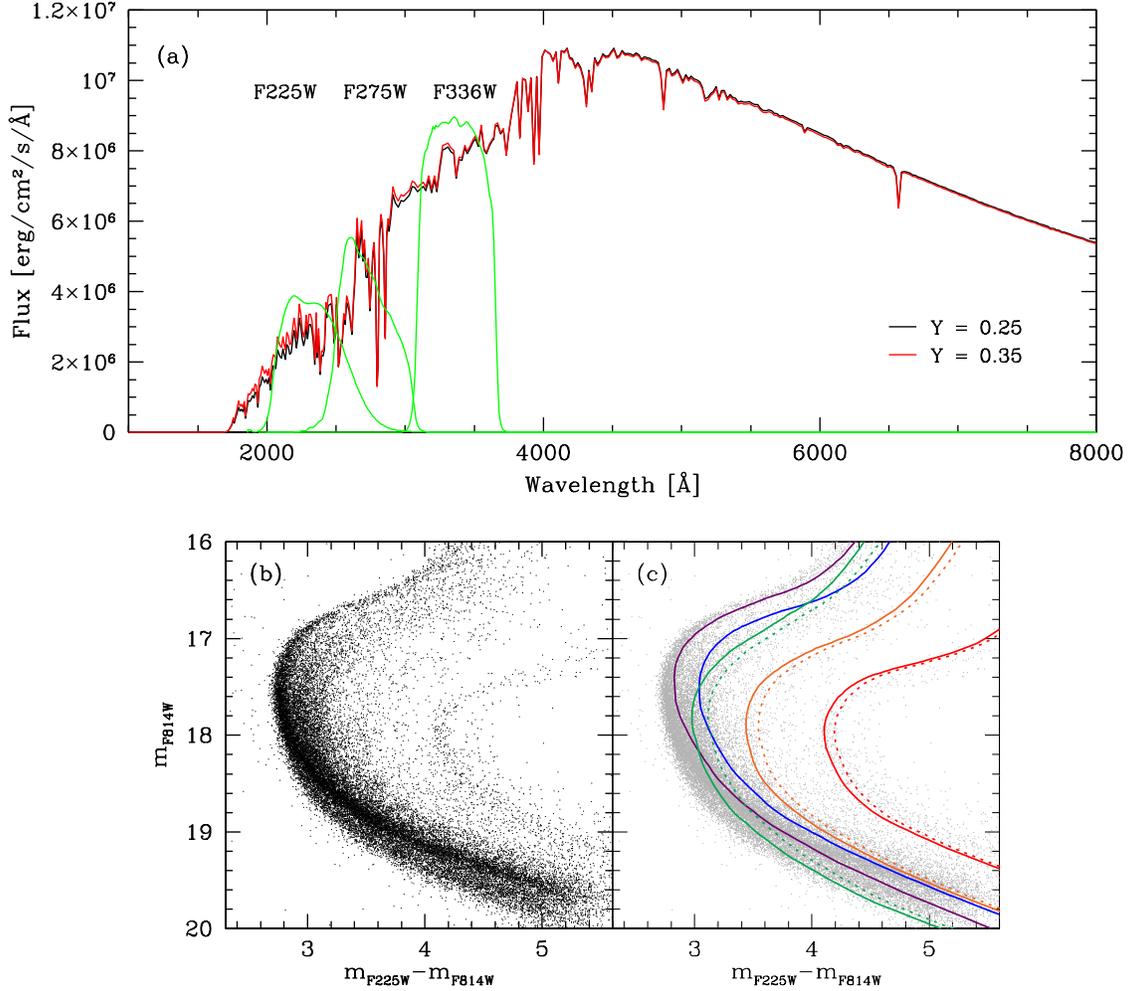}
\caption{
The effect of helium enhancement on the stellar spectra.
(a) Synthetic spectra \citep{cas03} of two stars near the MSTO, with the same $T \rm_{eff}$ = 6000 K,
log $g$ = 4.5, and [Fe/H] = $-$1.5, but different helium abundance (Y = 0.25 \& 0.35). Green lines are filter throughputs
for three UV-passbands (F225W, F275W, and F336W) of $HST$ WFC3/UVIS.
(b) Observed CMD of $\omega$~Cen in F225W and ACS/WFC F814W passbands (data from \citealt{bel10}).
(c) The effect of helium-enhanced spectra on the CMD. Color codings are as in Fig. 5. The dotted lines are
models from the normal helium (Y = 0.25) spectra, while solid lines are those from the helium-enhanced
(Y = 0.38$-$0.41) spectra. Adopted distant modulus and reddening are ($m$$-$$M)_{F814W}$ = 13.9 and
$E(F225W$$-$$F814W$) = 1.26, respectively.
}
\end{figure*}


\clearpage
\begin{figure}
\epsscale{0.45}
\plotone{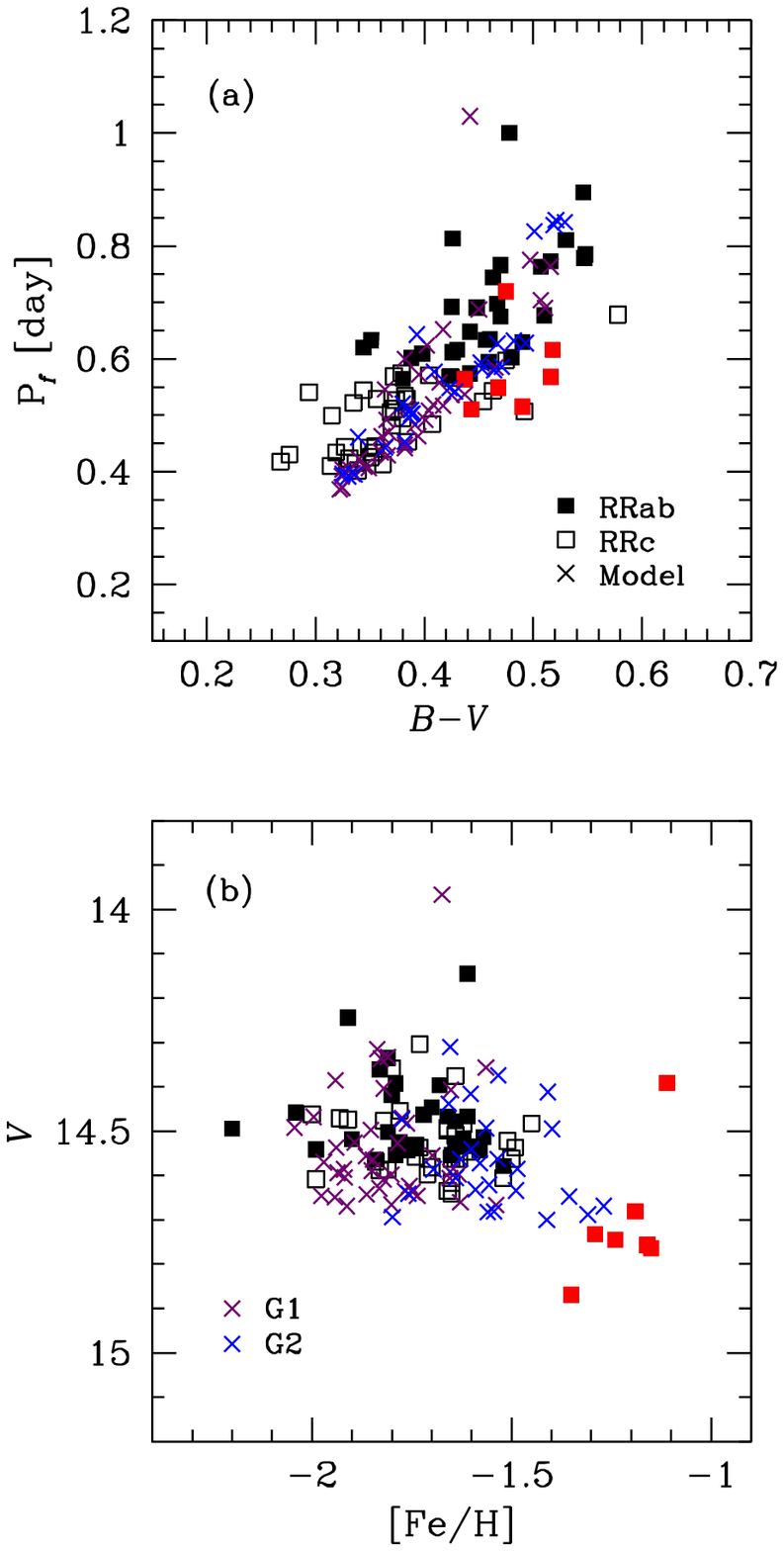}
\caption{
Comparison of the observed RR Lyrae variables (squares) in $\omega$~Cen with the synthetic HB models (crosses).
Red squares are metal-rich RRab stars ([Fe/H] $>$ $-$1.4). RR Lyrae data from \citet{sol06} and \citet{kal04}.
Purple and blue crosses are model RR Lyrae stars from G1 and G2, respectively.
This suggests that there is an additional minor subpopulation between G2 and G3, having normal helium-abundance
and [Fe/H] $\approx$ $-$1.3. Adopted distant modulus and reddening are ($m$$-$$M)_{V}$ = 14.1 and
$E(B$$-$$V$) = 0.14, respectively (see text).
}
\end{figure}


\clearpage
\begin{figure*}
\epsscale{0.9}
\plotone{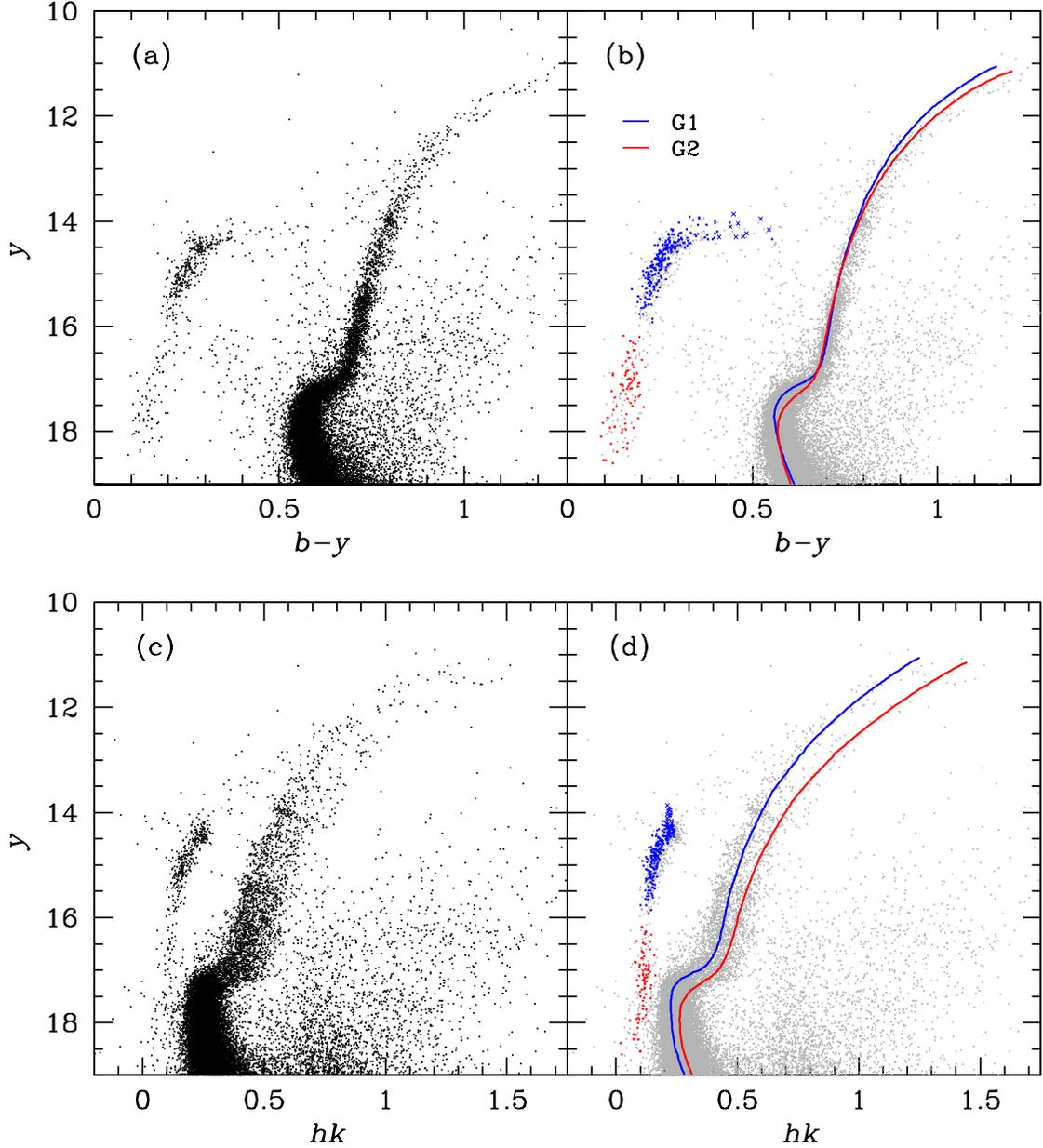}
\caption{
Comparison of our models with the observations for M22.
(a), (c) Str\"{o}mgren $b$, $y$, and calcium ($Ca$) narrow-band photometry from S.-I. Han et al. (2012, in preparation),
where the $hk$ index is defined as $hk$ = $(Ca-b)-(b-y)$.
(b), (d) Our population models compared on the observed CMDs.
Adopted distant modulus and reddening are ($m$$-$$M)_{y}$ = 13.80, $E(b$$-$$y$) = 0.255, and $E(hk)$ = $-$0.012.
}
\end{figure*}

\clearpage
\begin{figure*}
\epsscale{1.0}
\plotone{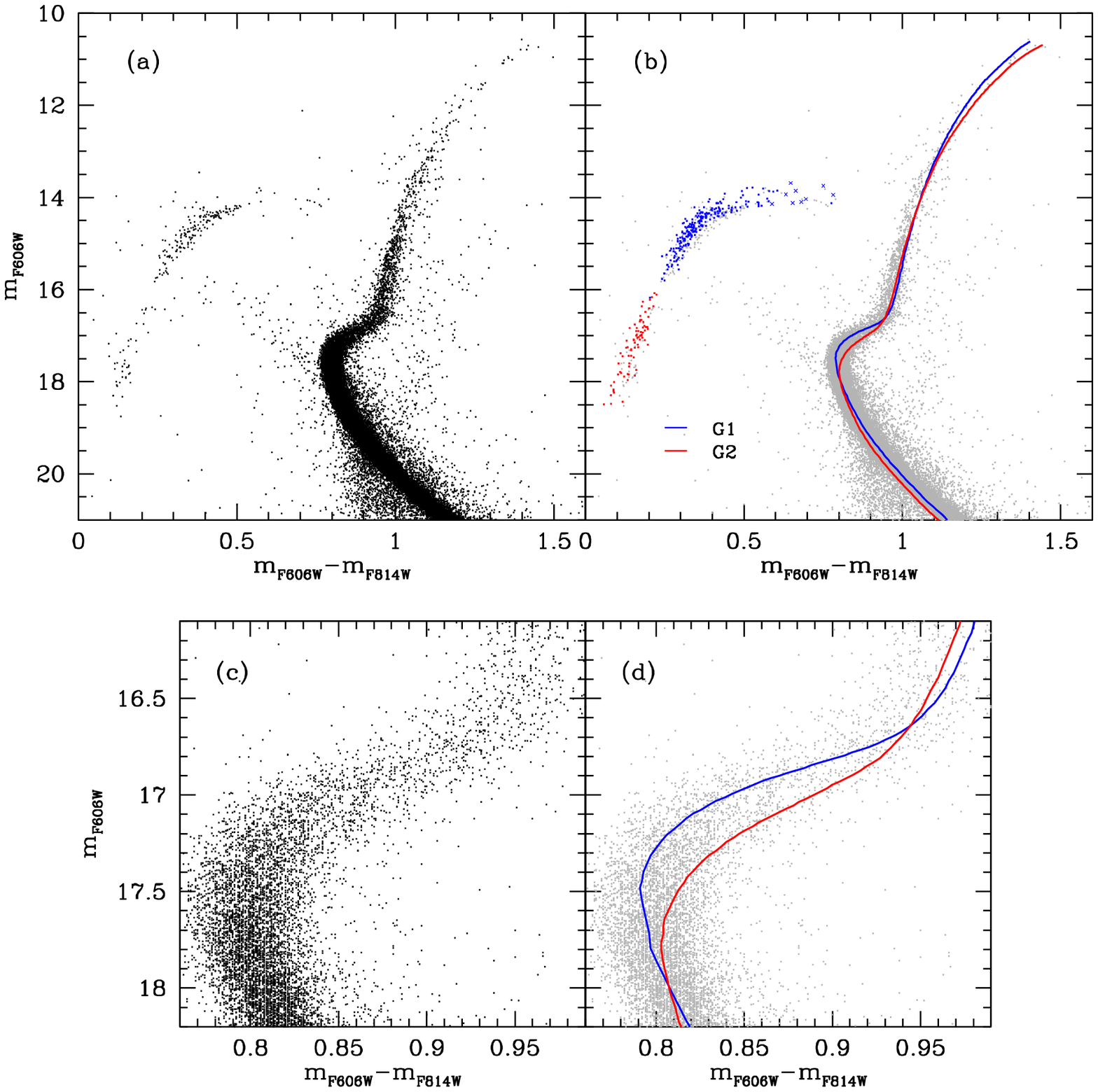}
\caption{
Comparison of our models with the observations for M22.
(a) Observed CMD by $HST$ ACS/WFC for the F606W and F814W passbands. Photometric data from the ACS survey \citep{sar07,and08}.
(b) Our models compared on the observed CMD.
(c), (d) Same as (a) \& (b) but zoomed in for the SGB region.
A clearer view of the SGB split can be seen in \citet[][see their Fig. 2]{mil10}.
Adopted distant modulus and reddening are ($m$$-$$M)_{F606W}$ = 13.69 and $E(F606W$$-$$F814W$) = 0.355, respectively.
}
\end{figure*}

\clearpage
\begin{figure}
\epsscale{0.7}
\plotone{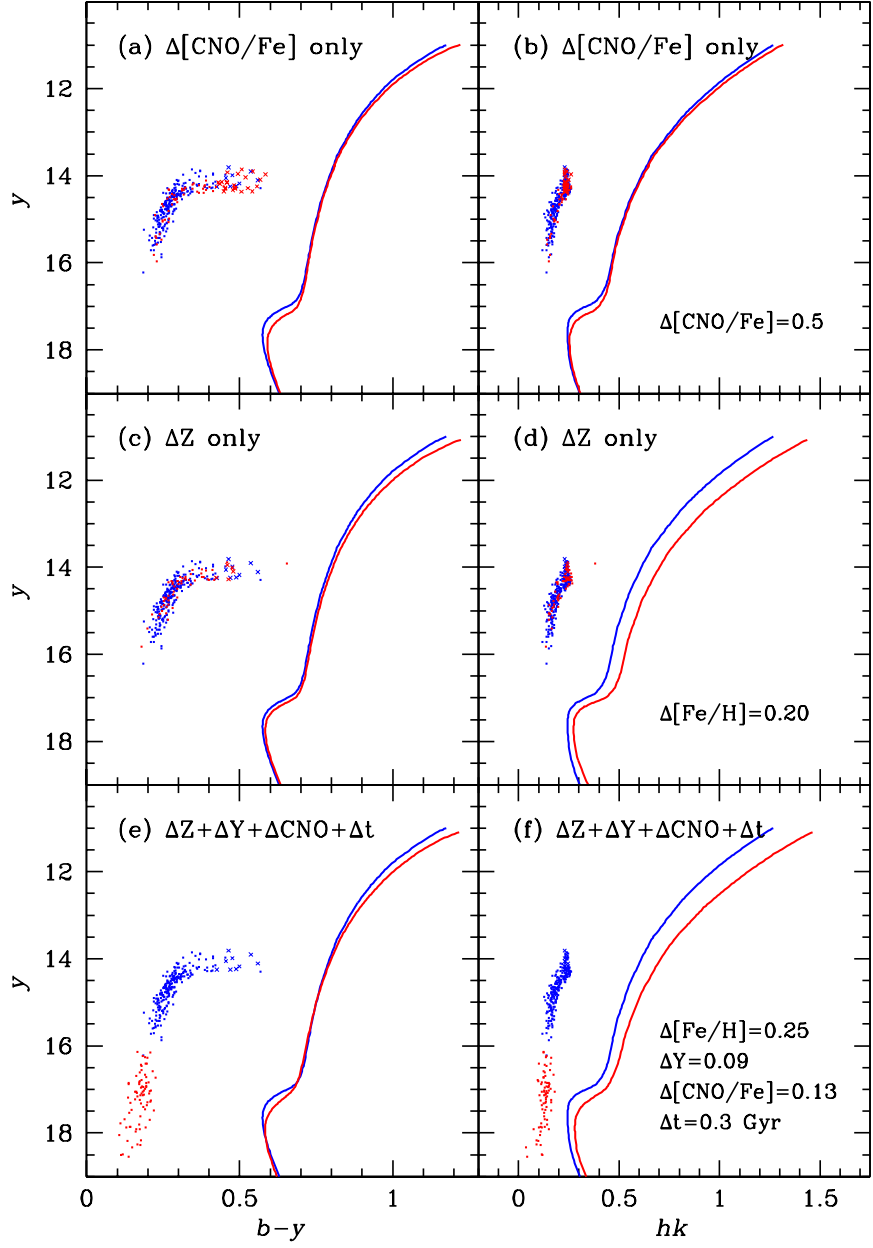}
\caption{
The effects of variations in the total CNO abundance, metallicity, and helium abundance in our models for M22
(see text).
}
\end{figure}

\clearpage
\begin{figure*}
\epsscale{0.9}
\plotone{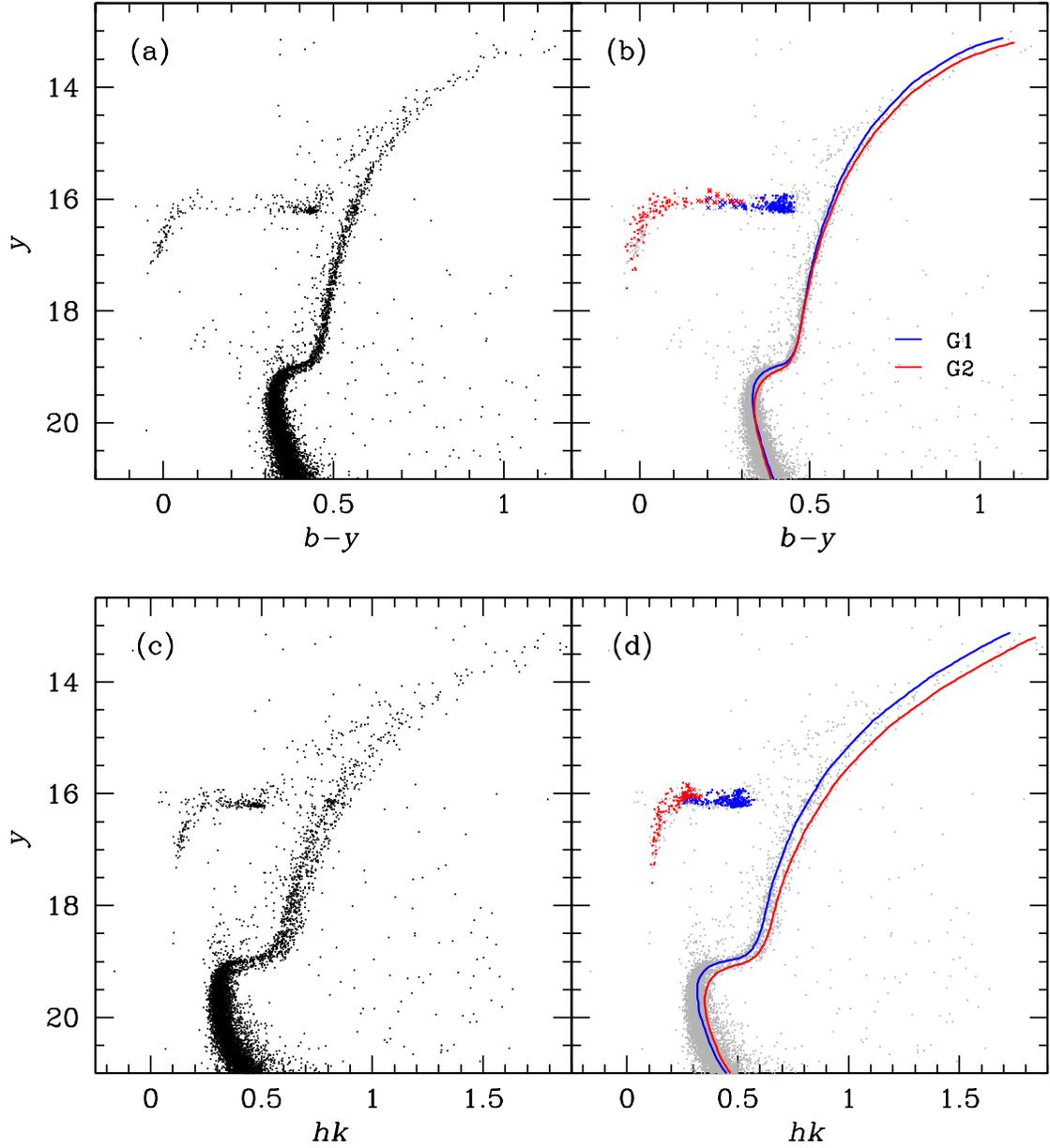}
\caption{
Same as Fig. 11, but for NGC~1851 (data from S.-I. Han et al. 2012, in preparation).
Adopted distant modulus and reddening are ($m$$-$$M)_{y}$ = 15.62, $E(b$$-$$y$) = 0.015, and $E(hk)$ = $-$0.015.
}
\end{figure*}

\clearpage
\begin{figure*}
\epsscale{1.0}
\plotone{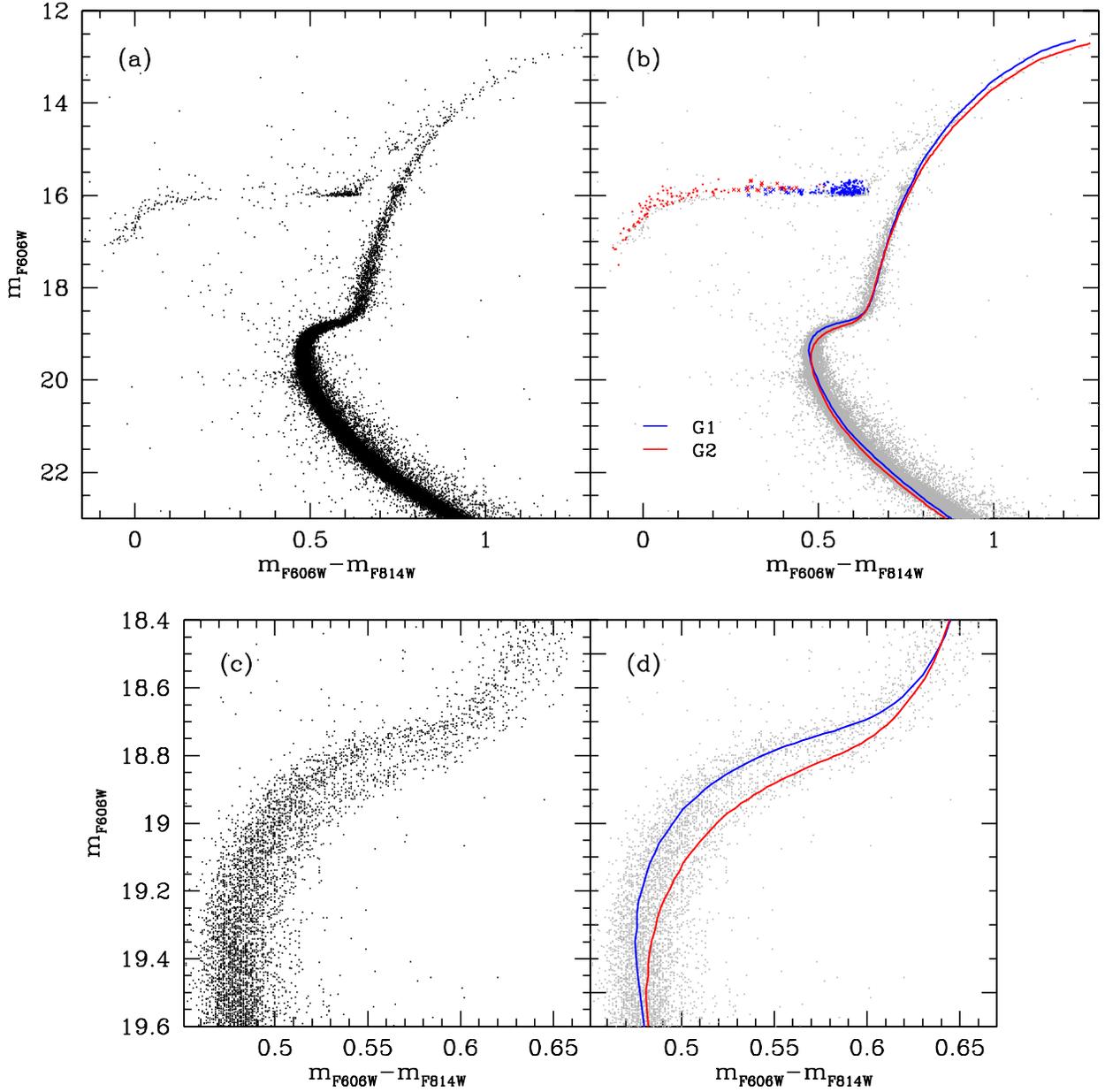}
\caption{
Same as Fig. 12, but for NGC~1851. Data from the ACS survey \citep{mil08}.
A clearer view of the SGB split can be seen in \citet[][see their Fig. 2]{mil10}.
Adopted distant modulus and reddening are ($m$$-$$M)_{F606W}$ = 15.52 and $E(F606W$$-$$F814W$) = 0.025, respectively.
}
\end{figure*}

\clearpage
\begin{figure}
\epsscale{0.65}
\plotone{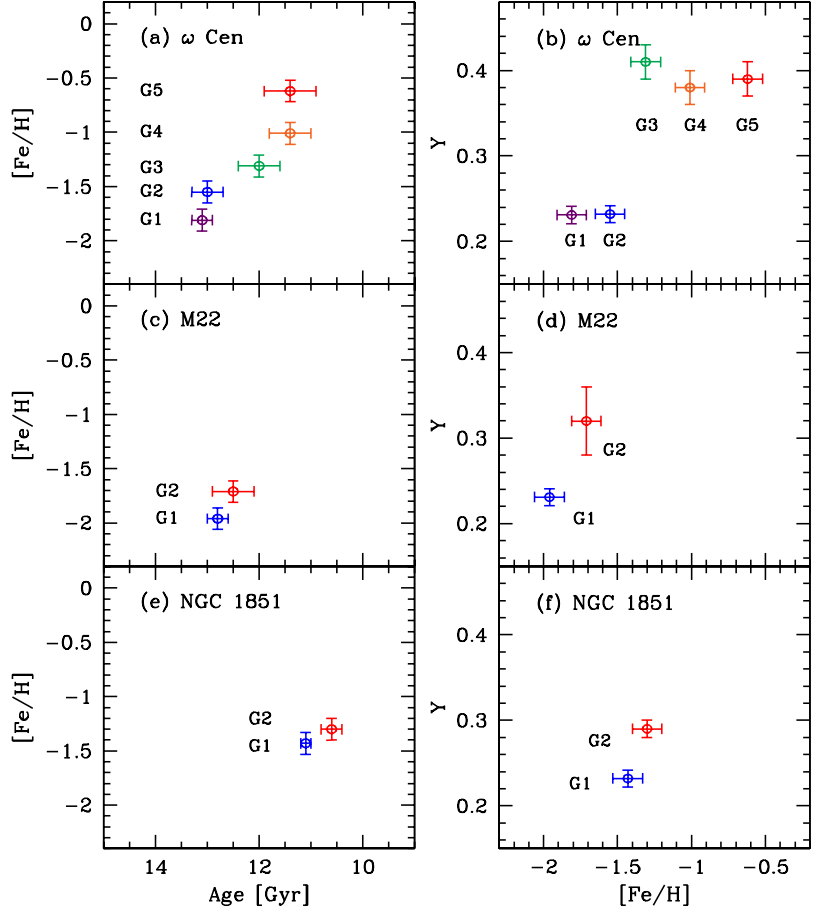}
\caption{
Schematic illustrations for the star formation histories and accompanying enrichments of metallicity
and helium abundance in three GCs considered in this paper.
The formation time scales of all subpopulations appear to be $\sim$1.7 Gyr for $\omega$~Cen and less
than 1 Gyr for M22 and NGC~1851. Metal-rich subpopulations in these GCs are also enhanced in helium abundance.
There is a dramatic increase of helium abundance between G2 and G3 in $\omega$~Cen.
One sigma error bars for age and helium abundance are from our $\chi^2$ minimization,
while the uncertainty in [Fe/H] is assumed to be 0.1 dex.
}
\end{figure}

\end{document}